\documentstyle[12pt,epsf]{article}

\newcommand{\CC}{\mbox{${\rm \:  C\!\!\! I
\;\;}$}}

\newcommand{\RR}{\mbox{${\rm \:  R\!\!\!\! I
\;\;}$}}

\newcommand{\vs}{\vspace{0.25cm}}

\newtheorem{theorem}{Theorem}
\newtheorem{itlemma}{Lemma}[section]
\newtheorem{itproposition}[itlemma]{Proposition}
\newtheorem{itcorollary}[itlemma]{Corollary}
\newtheorem{itremark}[itlemma]{Remark}
\newtheorem{itremarks}[itlemma]{Remarks}
\newtheorem{itdefinition}[itlemma]{Definition}
\newtheorem{itexample}[itlemma]{Example}

\newenvironment{lemma}{\begin{itlemma}\rm}{\end{itlemma}} 
\newenvironment{remark}{\begin{itremark}\rm}{\end{itremark}} 
\newenvironment{remarks}{\begin{itremarks} \rm}{\end{itremarks}}
\newenvironment{corollary}{\begin{itcorollary}\rm}{\end{itcorollary}}
\newenvironment{proposition}{\begin{itproposition}\rm}{\end{itproposition}}
\newenvironment{definition}{\begin{itdefinition}\rm}{\end{itdefinition}}
\newenvironment{example}{\begin{itexample}\rm}{\end{itexample}}
\newenvironment{fact}{\noindent {\em Fact}. \ \ }{\hfill \medskip}
\newenvironment{proof}{\noindent {\em Proof}.\ \
}{\hspace*{\fill}$\Box$\medskip}
\newenvironment{claim}{\noindent {\em Claim}. \ \ }{\hfill \medskip}
\newcommand{\be}[1]{\begin{equation}\label{#1}}
\newcommand{\ee}{\end{equation}}
\newcommand{\bl}[1]{\begin{lemma}\label{#1}}
\newcommand{\br}[1]{\begin{remark}\label{#1}}
\newcommand{\brs}[1]{\begin{remarks}\label{#1}}
\newcommand{\bt}[1]{\begin{theorem}\label{#1}}
\newcommand{\bd}[1]{\begin{definition}\label{#1}}
\newcommand{\bp}[1]{\begin{proposition}\label{#1}}
\newcommand{\bc}[1]{\begin{corollary}\label{#1}}
\newcommand{\bfact}[1]{\begin{fact}\label{#1}}
\newcommand{\bex}[1]{\begin{example}\label{#1}}
\newcommand{\ec}{\end{corollary}}
\newcommand{\efact}{\end{fact}}
\newcommand{\eex}{\end{example}}
\newcommand{\el}{\end{lemma}}
\newcommand{\er}{\end{remark}}
\newcommand{\ers}{\end{remarks}}
\newcommand{\et}{\end{theorem}}
\newcommand{\ed}{\end{definition}}
\newcommand{\ep}{\end{proposition}}
\newcommand{\epr}{\end{proof}}
\newcommand{\bpr}{\begin{proof}}
\newcommand{\bcl}{\begin{claim}}
\newcommand{\ecl}{\end{claim}}

\newcommand{\bi}{\begin{itemize}}
\newcommand{\ei}{\end{itemize}}
\newcommand{\ben}{\begin{enumerate}}
\newcommand{\een}{\end{enumerate}}
\newcommand{\text}[1]{\hbox{\rm \ #1\ \/}}

\title{Lie Algebraic Analysis and Control of Quantum
Dynamics}

\author{Domenico D'Alessandro\thanks{Department of Mathematics, Iowa State
University, Ames, Iowa, U.S.A.\ \ Electronic address:
daless@iastate.edu. This work was supported by NSF, under CAREER
Grant ECS0237925 and Grant ECCS0824085. }}

\begin{document}

\maketitle



\begin{abstract}
In this paper, we show how to use the analysis of the Lie algebra
associated with  a quantum mechanical system to study its dynamics
and facilitate the design of controls. We give algorithms to
decompose the dynamics and describe their application to the control
of two coupled spin $\frac{1}{2}$'s.

\end{abstract}

\section{Introduction}
\label{intro} For several quantum mechanical systems subject to a
control action, an appropriate model is the {\it Schrodinger
operator (matrix)  equation}
 \be{Mgensys}
\dot X=-iH(u(t))X, \qquad X(0)={\bf 1}.
 \ee
In this equation, $H$ is called the {\it Hamiltonian} and it is an
Hermitian matrix function of the control $u\equiv u(t)$
and $X=X(t)$ is a unitary matrix for every time $t$, with $\bf 1$
denoting the identity. Equation (\ref{Mgensys}) is an appropriate
model for many quantum phenomena under three main assumptions: 1)
The quantum mechanical system under consideration can be adequately
approximated as a system having a finite number of energy levels;
2) The interaction with the external environment (decoherence) is
negligible; 3) The control, which usually represents an
appropriately shaped electro-magnetic field, can be treated as a
classical field (semiclassical approximation). Many physical systems
share a model of the form (\ref{Mgensys}). Examples are given by
systems of particles with spin subject to a control magnetic field
such as in NMR and EPR, molecular systems where the control is an
electric field and several implementations of quantum information
processing. Here the control can be seen as an action allowing us to
switch among different Hamiltonians to implement given quantum
evolutions (quantum gates). The books \cite{BUKSam},
\cite{MikoBook}, \cite{Zhao}  present physical examples of systems
sharing the model (\ref{Mgensys}).

 It is well known (see, e.g, \cite{MikoBook}, \cite{sussJ})   that
 the set of operators $X$ reachable for (\ref{Mgensys}) by varying the control $u$ is the
connected Lie subgroup of $U(n)$ corresponding to the Lie algebra
$\cal L$ generated by the set $ {\cal F}:=\{iH(u)|u \in {\cal U}
\},$ that is, the smallest Lie subalgebra of $u(n)$ containing
${\cal F}$ (see the Appendix). The Lie algebra $\cal L$ is called
the {\it dynamical Lie algebra} associated with the system and the
associated connected Lie subgroup of $U(n)$ will be denoted here by
$e^{\cal L}$.\footnote{Extending this notation, we shall use the
notation $e^{\cal K}$ for the connected Lie group associated with  a
Lie algebra ${\cal K}$. See the Appendix for definitions.} There
exists a very simple algorithm to calculate ${\cal L}$: One starts
with a basis of ${\cal F}$, $F_1,\ldots, F_r$. If $r=n^2$ or
$r=n^2-1$, one stops because ${\cal L}=u(n)$ or ${\cal L}=su(n)$,
respectively. In this case,  $e^{\cal L}=U(n)$ or $e^{\cal L}=SU(n)$
and the system is said to be controllable.
If this is not the case, one performs the Lie brackets of {\it depth
1} $[F_j, F_k]$, $j \not= k$ and select the ones that are linearly
independent together  with $\{F_1, \ldots, F_r\} $, say
$D_1,\ldots,D_s$, if any. Then one performs Lie brackets of depth
$2$ which are Lie brackets of $D_1,\ldots,D_s$ with the $F_j$'s,
$j=1, \ldots,r$ and select matrices that are linearly independent
together  with $F_1, \ldots, F_r,D_1,\ldots,D_s$. One goes on this
way until one does not find any new linearly independent matrices.
The set of matrices thus found is a basis of the dynamical Lie
algebra ${\cal L}$. If the size of this set is $n^2$ or $n^2-1$,  we
are in the controllable case. Otherwise the system is not
controllable.\footnote{See subsection 3.2.1 of \cite{MikoBook} for
further discussion of this procedure.} However the Lie algebra
${\cal L}$ gives us information about the nature of the dynamics as
we shall see next.

In section \ref{deco} we use the Levi's decomposition of Lie
algebras to obtain a decomposition of the dynamics for
(\ref{Mgensys}). We highlight the simplifications that follow from
the fact that ${\cal L}$ is a subalgebra of $u(n)$. Levi's
decomposition is a classical result in Lie algebra theory but its
impact in quantum control has not been considered
before.\footnote{One exception is the book \cite{MikoBook}. However
we shall go beyond what is in this book here by pointing out the
simplifications in this decomposition in the quantum case and giving
explicit algorithms for calculation of the decomposition.} In
section \ref{algo}, we give algorithms to calculate such a
decomposition. Algorithms for general Lie algebras are known
\cite{DeGraaf} but simplified algorithms can be given in the case of
interest here. In some cases, our algorithms will be a simplified
version of the ones in \cite{DeGraaf} in some cases  different
algorithms will be given. We shall point out this as we go on. In
section \ref{examples} we use this decomposition for a control
problem for two
 spin $\frac{1}{2}$ particles.


\section{Decomposition of Quantum  Dynamics}
\label{deco}

 Every
Lie algebra ${\cal L}$ over the field of reals $\RR$ is the
semidirect sum of a semisimple Lie algebra $\cal S$ and the  maximal
solvable ideal in ${\cal L}$,  ${\cal R}$, called the {\it
radical},\footnote{See definitions in the Appendix.} that is,
\be{Levidec} {\cal L} = {\cal S} \oplus {\cal R}. \ee Semidirect sum
means that \be{Levidec2} [{\cal S}, {\cal S}] \subseteq {\cal S},
\qquad [{\cal S}, {\cal R}] \subseteq {\cal R}, \qquad [{\cal R},
{\cal R}] \subseteq {\cal R}. \ee This is a classical result known
as the {\it Levi decomposition} (see, e.g., \cite{DeGraaf}). ${\cal
S}$ is called the {\it Levi subalgebra}. It is the direct sum of $p
\geq 1$ simple subalgebras ${\cal S}_j$, $j=1,\ldots,p$, i.e.,
\be{ssdec} {\cal S}= \bigoplus_{j=1}^p {\cal S}_j.  \ee Direct sum
means that $ [{\cal S}_l, {\cal S}_b]=\{0\},$ when $l \not= b$. In
our case, the fact that the dynamical Lie algebra ${\cal L}$ is a
subalgebra of $u(n)$ implies several important simplifications.
 \bt{MSimplyf}\footnote{This fact is mentioned in the paper \cite{Tannor} but
without a proof. We provide a proof here.} If ${\cal L} \subseteq
u(n)$ then the semidirect sum in (\ref{Levidec}) is a direct sum,
i.e., $[{\cal S}, {\cal R}]=\{ 0\},  $ and ${\cal R}$ is Abelian,
i.e., $[{\cal R}, {\cal R}]=\{0\} $. \et This theorem is a
consequence of Lie's theorem (cf., e.g., \cite{Humphreys}, Corollary
A in Section 4.1).

\bt{Lietheo} (Lie's Theorem) Let ${\cal R}_{\otimes \CC}$ be a
solvable Lie algebra of $n \times n$ matrices over the complex
field. Then there exists a change of coordinates (i.e., a similarity
transformation) to put all the elements in ${\cal R}_{\otimes \CC}$
in upper triangular form. \et

We now give the proof of Theorem \ref{MSimplyf}.

\bpr We first prove that ${\cal R}$ is Abelian. We shall consider
 the field extension (see the Appendix and, e.g., \cite{SagleWalde}
for a more in depth discussion) of ${\cal R}$ to the complex field,
${\cal R}_{\otimes \CC}$. We shall show that ${\cal R}_{\otimes
\CC}$ is Abelian and this implies ${\cal R}$ Abelian. In fact ${\cal
R}$ is solvable (Abelian) if and only if ${\cal R}_{\otimes \CC}$ is
solvable (Abelian). Since ${\cal R}_{\otimes \CC}$ is solvable,
according to Lie's Theorem \ref{Lietheo} it can be realized as upper
triangular matrices. As a consequence, it can be written as the sum
of two subalgebras: a nilpotent Lie algebra ${\cal N}$,
corresponding to strictly upper triangular matrices, and an Abelian
Lie algebra $\cal T$, corresponding to diagonal matrices in the
coordinates indicated in Lie's theorem. Moreover, ${\cal N}$ is an
ideal in ${\cal R}_{\CC}$. Now consider $R$ and $P$ in ${\cal N}$.
Since ${\cal N}$ is nilpotent, there exists a $k
>0 $ such that, $ad_R^kP:=[R,[R,[\cdots,[R,P]]]]=0$ where the Lie bracket is taken
$k$ times.
%
 Now,  since $R$ is skew-Hermitian, there is no loss of generality in
assuming that $R$ is diagonal, i.e., $R:=\text{diag}(i
\lambda_1,\ldots,i \lambda_n)$. Moreover we calculate  \be{op9}
({ad^k_RP})_{j,l}=p_{j,l}{i}^k(\lambda_j-\lambda_l)^k, \qquad
\forall j,l=1,...,n,  \ee where $i:=\sqrt{-1}$.  From this
expression it follows that if ${ad^k_XP}_{j,l}$ is zero for some
$k$,  it must be zero for {\it every} $k$ and in particular for
$k=1$. Therefore $R$ and $P$ commute and ${\cal N}$ is Abelian.
Consider now $N \in {\cal N}$ and $T \in {\cal T}$. Since ${\cal N}$
is an ideal $[N,T]$ is in ${\cal N}$ and therefore $[N,[N,T]]=0$
since ${\cal N}$ is Abelian. The calculation (\ref{op9}) shows that
$N$ and $T$ commute and ${\cal R}_{\otimes \CC}$ is the  sum of two
commuting Abelian subalgebras and it is therefore Abelian.

The proof that $[{\cal S}, {\cal R}]=0$ use the same calculation
(\ref{op9}). Since ${\cal R}$ is an Abelian ideal (cf.
(\ref{Levidec2})), $ad_R^2P=0$ for every $R\in {\cal R}$ and $P \in
{\cal S}$, which, from (\ref{op9}), implies $ad_RP=0$. \epr

This decomposition of the dynamical Lie algebra ${\cal L}$ has
immediate consequences for the Lie group of possible evolutions
$e^{\cal L}$ and for the dynamics of the quantum system
(\ref{Mgensys}). For every control $u$, the solution of
(\ref{Mgensys}) $X=X(t)$ factorizes as \be{factoriz2}
X(t)=R\prod_{j=1}^pS_j. \ee Here $S_j \in e^{{\cal S}_j}$ and $R \in
e^{\cal R}$ and all the factors in (\ref{factoriz2}) commute.
Moreover $R$ is itself the product of elements belonging to one
dimensional subgroups. Write ${\cal R}$ as the sum of one
dimensional Lie algebras ${\cal R}= \bigoplus_{l=1}^q {\cal R}_l$,
then $R=\prod_{l=1}^q R_l$, with $R_l \in e^{{\cal R}_l}$.
Controlling the system (\ref{Mgensys}) means controlling in parallel
the systems $\dot S_j=-iH_{Sj}(u)S_j, \qquad S_j(0)={\bf 1}, \qquad
j=1,\ldots p, $ and $
 \dot R_l=-iH_{Rl}(u)R_l,
\qquad R_l(0)={ \bf 1}, \qquad l=1,\ldots,q, $ where $-iH_{Sj}$ and
$-iH_{Rl}$ are the components of $-iH(u)$ in ${\cal S}_j$,
$j=1,...,p$, and ${\cal R}_l$, $l=1,\ldots,q$, respectively.

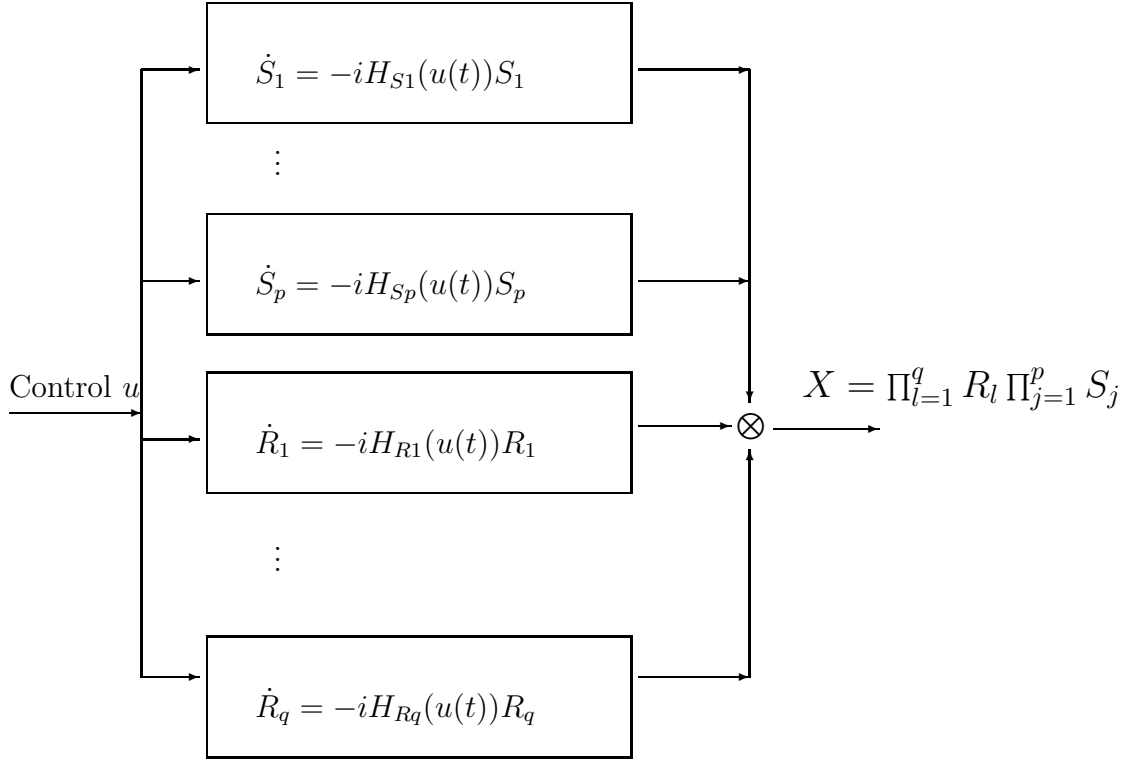
\begin{figure}[ht]

\begin{picture}(340,300)(0,-20)

\linethickness{0.6pt}

\put(60,120){\vector(1,0){50}}

\put(60,126){{Control $u$}}

\put(135,230){\framebox(160,45)}

\put(135,150){\framebox(160,45)}

\put(160,210){$\vdots$}

\put(135,90){\framebox(160,45)}

\put(110,120){\line(0,1){130}}

\put(110,250){\line(1,0){10}}

\put(120,250){\vector(1,0){12}}

\put(110,170){\line(1,0){10}}

\put(120,170){\vector(1,0){12}}

\put(160,60){$\vdots$}

\put(110,120){\line(0,-1){100}}

\put(110,110){\vector(1,0){22}}
\put(298,115){\vector(1,0){35}}


\put(298,20){\vector(1,0){42}}

\put(110,20){\vector(1,0){22}}

\put(340,20){\line(0,1){50}}

\put(298,170){\vector(1,0){42}}

\put(298,250){\vector(1,0){42}}

\put(340,250){\vector(0,-1){126}}

\put(340,70){\vector(0,1){36}}

 \put(335,112){{${ \bigotimes}$}}

 \put(349,114){\vector(1,0){40}}


\put(153,245){$\dot S_1=-iH_{S1}(u(t))S_1$}


\put(153,165){$\dot S_p=-iH_{Sp}(u(t))S_p$}

\put(153,105){$\dot R_1= -iH_{R1}(u(t))R_1$ }


\put(153,5){$\dot R_q= -iH_{Rq}(u(t))R_q$ }

\put(135,-10){\framebox(160,45)}


\put(360,126){\large{{$X=\prod_{l=1}^q R_l  \prod_{j=1}^p S_j$}}}

\end{picture}

\caption{Structure of a quantum control system. The control $u$
drives simultaneously $p+q$ systems on Lie groups (which are simple
Lie groups or one-dimensional Lie groups). The total evolution $X$
is the commuting product of the evolutions on the various
subgroups.} \label{Figure2}
\end{figure}

Every, finite dimensional, quantum system (\ref{Mgensys}) has the
structure of $p+q$ subsystems in parallel of Figure \ref{Figure2}.
The first $p$ subsystems vary on simple Lie groups for which a
classification is known \cite{tables}, \cite{Knapp}. The remaining
$q$ subsystems vary on one dimensional Lie groups. The total
evolution is the commuting product of the evolutions on the various
 subgroups. To obtain the decomposition of the dynamics, we need to
find bases for the subalgebras, ${\cal S}_j$ and ${\cal R}_l$,
$j=1,\ldots,p$, $l=1,\ldots q$, of ${\cal L}$ from a basis of ${\cal
L}$. Next, we give algorithms for this task.

\section{Algorithms}
\label{algo}

The book  \cite{DeGraaf} contains  several algorithms for general
Lie algebras.
In our case, we only need to consider are subalgebras of $u(n)$.
This allows us in some cases to give new and simple algorithms and
in other cases to simplify the algorithms of \cite{DeGraaf}.

Consider a basis $\{L_1,\ldots,L_s\}$ of $\cal L$. The calculation
of bases of the two subspaces ${\cal S}$ and ${\cal R}$ in
(\ref{Levidec}) is very simple in the case of interest here. It is
easily seen that ${\cal R}$ is the center of ${\cal L}$, and
therefore it is the space of the solutions of the system of
$s=\dim{\cal L}$ equations $ [R,L_j]=0, \qquad j=1,\ldots,s, $ in
the variable $R\in {\cal L}$. Moreover, using the fact that for
every semisimple Lie algebra ${\cal S}$, ${\cal S}=[{\cal S}, {\cal
S}]$, along with $[{\cal S}, {\cal R}]=0$ and that ${\cal R}$ is
Abelian, we have \be{Msimpcalc} [{\cal L}, {\cal L}] =[{\cal S}
\oplus {\cal R}, {\cal S} \oplus {\cal R}]=[{\cal S}, {\cal S}]=
{\cal S}.  \ee Therefore the set of $\pmatrix{s \cr 2}$ matrices,
$[L_j, L_k]$, $j \not= k$, spans ${\cal S}$.

The algorithm of \cite{DeGraaf}  to find the solvable radical (cf.
Section 2.6 in \cite{DeGraaf}) first finds the product space $[{\cal
L},{\cal L}]$ and then the adjoint representations of all elements
in a basis of ${\cal L}$. Then the algorithm solves a linear system
of $\dim [{\cal L}, {\cal L}]$ equations in $\dim {\cal L}$ unknowns
obtained by using the Killing form (cf. the Appendix). Then one goes
on to calculate the Levi subalgebra (cf. Section 4.13 in
\cite{DeGraaf}). In our case, we have  avoided the calculation and
storage of the adjoint representation. Moreover,
 once one knows $[{\cal L}, {\cal L}]$ there is no need to apply any
algorithm to find the Levi's subalgebra ${\cal S}$ as we have
(\ref{Msimpcalc}).

The calculation of the simple ideals of ${\cal S}$, ${\cal S}_j$,
$j=1,\ldots,p$,  is more complicated. We follow the path indicated
in \cite{DeGraaf}. A preliminary step is the calculation of the
so-called {\it primary decomposition} of ${\cal S}$ which is also of
interest to understand the structure of ${\cal S}$.

\subsection{Calculation of the primary decomposition of ${\cal S}$}

The following definition is of interest for general Lie algebras.
\bd{Cartansubd} A Cartan subalgebra of a Lie algebra ${\cal S}$ is a
nilpotent subalgebra ${\cal A}$ which is equal to its {\it
normalizer}, that is $ {\cal A}=\{S \in {\cal S}|[S, {\cal A}]
\subseteq {\cal A} \}.$ \ed In the case of ${\cal S} \subseteq
u(n)$, we have the following \bp{nilab} Every nilpotent subalgebra
of $u(n)$ is Abelian. In particular the Cartan subalgebra of ${\cal
S}$ is Abelian.  \ep \bpr The proof uses the same calculation
(\ref{op9}) and argument in Theorem \ref{MSimplyf} to show that if
$ad_S^kX=0$, for some $k$, and two elements $S$ and $X$ in the Lie
algebra, then it must be $ad_SX=0$, i.e., $S$ and $X$ commute. \epr

The following algorithm calculates the Cartan subalgebra for ${\cal
S}$ semisimple and ${\cal S} \subseteq u(n)$.

\verb"Algorithm 1"

\begin{enumerate}

\item Given the semisimple Lie algebra ${\cal S}$, set   ${\cal A}=\{0\}$.

\item Select an element $X \not= 0 $ in ${\cal S}$.

\item Calculate the set of elements in $\cal S$ which commute with
$X$. Call this set $\cal D$.

Notice ${\cal D}$ is also a subalgebra of $u(n)$.\footnote{The fact
that it is a Lie algebra, i.e., $[D_1,D_2] \in {\cal D}$, for $D_1,
D_2 \in {\cal D}$ follows immediately from an application of the
Jacobi identity. We have $[[D_1,D_2],X]=-[[D_2,X],D_1]-[[X,D_1],
D_2]=0$.} Therefore (just like ${\cal L}$ above), it has a Levi
decomposition in its semisimple part which is equal to $[{\cal D},
{\cal D}]$ and the center, $C({\cal D})$. This justifies the next
step.

\item Write ${\cal D}=[{\cal D}, {\cal D}] \oplus C({\cal D})$ where
$C({\cal D})$ is the center of ${\cal D}$.

\item Set ${\cal A}={\cal A} \oplus C({\cal D})$.

\item If $[{\cal D}, {\cal D}] =0$ Stop and return ${\cal A}$ as
Cartan subalgebra, otherwise set ${\cal S}=[{\cal D}, {\cal D}]$ and
go to step 2.

\end{enumerate}

The algorithm converges because at each step ${\cal S}$ is
semisimple and $\cal D$ is a proper subspace of ${\cal S}$,
otherwise ${\cal S}$ would have an element which commutes with all
of ${\cal S}$ which contradicts semisimplicity. To show that the
algorithm gives in fact a  Cartan subalgebra of ${\cal S}$ we have
to show two facts

1) The resulting ${\cal A}$ is nilpotent (it is in fact Abelian).

2) Every $S \in {\cal S}$ which is such that $[S, {\cal A}]
\subseteq {\cal A}$ is an element of ${\cal A}$.

\bpr (Proof of 1) and 2) above) Let us denote by ${\cal A}_k$ and
${\cal D}_k$ the Lie algebras ${\cal A}$ and ${\cal D}$ obtained
after the $k$-th step is complete.
 We show 1) by induction on the steps of
the algorithm.  ${\cal A}_1$ is definitely Abelian as it is the
center of ${\cal D}_1$. Moreover it commutes with ${\cal D}_1$. The
inductive step shows that these two properties are true at each
step.  Assume they are  true at step $k-1$. At step $k$, ${\cal
A}_k={\cal A}_{k-1} \oplus { C}({\cal D}_k)$. However both ${\cal
A}_{k-1}$ and ${\cal C}({\cal D}_k)$ are Abelian and $[{\cal
A}_{k-1}, {\cal C}({\cal D}_k)]=\{0\}$, because ${\cal D}_k \subset
{\cal D}_{k-1}$, and we know by the inductive assumption that ${\cal
A}_{k-1}$ commutes with ${\cal D}_{k-1}$. This proves that ${\cal
A}_k$ is Abelian. Using again ${\cal A}_k={\cal A}_{k-1} \oplus {
C}({\cal D}_k)$, and the fact that ${\cal D}_k \subset {\cal
D}_{k-1}$, $[{\cal A}_k, {\cal D}_k]=\{ 0\}$ and the inductive step
is complete.

To show 2) assume $[S, {\cal A}] \subseteq {\cal A}$.  Then
repeating the argument following (\ref{op9}) $[S, {\cal A}]=\{ 0
\}$. Consider $X$ and ${\cal D}_1$ at the first step. Write
$S=S_1+S_0$ with $S_0 \in {\cal D}_1$ and $S_1$ in a complement of
${\cal D}_1$ in ${\cal S}$. Since $X \in {\cal A}$ and $S$ commutes
with ${\cal A}$, then $S$ commutes with $X$. Since $S_0$ commutes
with $X$ so does $S_1$. So, $S_1$ is in ${\cal D}_1$ and in a
complement of ${\cal D}_1$ which implies $S_1=0$. Therefore  $S \in
{\cal D}_1$ and we can write it as $S=D_1+A_1$ with $D_1 \in [{\cal
D}_1, {\cal D}_1]$ and $A_1 \in C({\cal D}_1)$ and therefore $\in
{\cal A}$. At the second step we pick another $X$ which turns out to
be again in ${\cal A}$. Since $S$ commutes with ${\cal A}$, ${ D}_1$
is in ${\cal D}_2$. We can then write $D_1=D_2+A_2$ with $D_2 \in
[{\cal D}_2,{\cal D}_2]$, and $A_2 \in {\cal A}$. Therefore
$S=D_2+A_2+A_1$. Proceeding this way, if the procedure ends after
$r$ steps,  we have that $[{\cal D}_r,{\cal D}_r]=\{0 \}$ and $S$
has the form $S=A_r+A_{r-1}+ \cdots + A_1$, with $A_j \in {\cal A}$,
for every $j=1,...,r$. Therefore $S \in {\cal A}$. \epr

\br{novelty} We remark that the above algorithm and proof is not
derived as a special case of the corresponding algorithm in
\cite{DeGraaf} (cf. Section 3.2 in \cite{DeGraaf}) but it is, to the
best of the author's knowledge, new. It gives the Cartan subalgebra
for the specific situation of interest in quantum control. It is
simpler than the general algorithm both because it involves fewer
notions of Lie algebra theory and because it involves fewer
operations. \footnote{The algorithm of \cite{DeGraaf} involves,
among the other things, the tuning a parameter so that a certain
vector space has a given dimension.}
\er

 The following definition refers to a general Lie algebra over a
general field (cf. \cite{DeGraaf} (Definitions 3.1.1 and 3.1.9)).

\bd{CPD} A {\it collected primary decomposition} of a semisimple Lie
algebra ${\cal S}$ with respect to a Cartan subalgebra ${\cal A}$ is
a vector space decomposition of the form \be{ioplm} {\cal S}:={\cal
A} \oplus {\cal V}_1 \oplus {\cal V}_2 \oplus \cdots {\cal V}_r, \ee
where

\begin{enumerate}
\item The subspaces ${\cal V}_j$'s, $j=1,\ldots,r$, are invariant
under $ad_X$, for every $X \in {\cal A}$, that is
$$
[{\cal A}, {\cal V}_j] \subseteq {\cal V}_j, \qquad j=1, \ldots, r.
$$
\item For every $X \in {\cal A}$,  and every  ${\cal
V}_j$, the minimum polynomial of $ad_X$ restricted to ${\cal V}_j$
is the power of an irreducible polynomial.

\item For any two subspaces ${\cal V}_j$ and ${\cal V}_k$, there
exists an $X \in {\cal A}$ such that the minimum polynomials of
$ad_X$ restricted to ${\cal V}_j$ and ${\cal V}_k$ are powers of two
{\it  different} irreducible polynomials.

\end{enumerate}

 Given ${\cal A}$ such a decomposition exists and is unique
(Theorem 3.1.10 of \cite{DeGraaf}). \ed

In our case, for every $X$,  the minimum polynomial  of $ad_X$
restricted to ${\cal V}_j$ must be of the type $(\lambda^2+a^2)$,
otherwise $ad_X$ would have eigenvalues with nonzero real parts
and-or eigenvalues with geometric multiplicity greater than one.
This is not possible because we have (in an appropriate basis)
$ad_X^T=-ad_X$ (see Proposition \ref{Richard} in Appendix).

An algorithm to calculate the collected primary decomposition is
given below. This algorithm was derived by applying the general
algorithm presented in \cite{DeGraaf} to the case considered here
(cf. Section 4.11 in \cite{DeGraaf}).

\vspace{0.25cm}

\verb"Algorithm 2"

\begin{enumerate}

\item Select an element $X \in {\cal A}$ such that $ad_X$ has $\dim
{\cal L} -\dim {\cal A}+1$ different eigenvalues. That is, except
for the $0$ eigenvalue (which has eigenspace equal to ${\cal A}$),
$ad_X$ is non-degenerate.

Such elements are called {\it splitting elements} and they exist
(Corollary 4.11.3 of \cite{DeGraaf}). To find such an $X$, notice
that, if $\{ A_1,\ldots,A_m\}$ is a basis of ${\cal A}$ then
$ad_{A_1},\ldots,ad_{A_m}$ all commute and they can be
simultaneously diagonalized. It is easier then to select
 real coefficients $c_j$, such that
 $\sum_{j=1}^m c_j ad_{A_j}$ has the desired property,  and
 $X=\sum_{j=1}^m c_j {A_j}$. For
higher dimensional problems it may be more convenient to use
randomized algorithms.

Let $X$ be the selected element, the minimum polynomial is of the
form $ m_{ad_X}(\lambda)=\prod_{j=1}^f (\lambda^2 + a_j^2), $ with
the $a_j$ all $\in \RR$, all different from each other and with one
of them equal to zero ($0$ is always an eigenvalue of $ad_X$, $X$
being an eigenvector). Moreover, from the choice of $X$ being
splitting $f=\frac{\dim{\cal L} - \dim{\cal A}}{2}+1$ and
$m_{ad_X}(\lambda)$ is equal to the characteristic polynomial except
(possibly) for the power of the monomial associated to the
eigenvalue $0$.\footnote{Notice in particular that the difference
between the dimension of ${\cal L}$ and that of ${\cal A}$ must be
an even number.}

\item Take as ${\cal V}_j$ the (two-dimensional) eigenspaces
associated with the pair of purely imaginary eigenvalues
corresponding to $a_j^2$. That is ${\cal V}_j=\{ V \in {\cal L} |
(ad_X^2+a_j^2 {\bf 1})V=0 \}, \qquad j=1,\ldots, \frac{\dim{\cal
S}-\dim{\cal A}}{2}.$

\end{enumerate}

We now prove that the decomposition obtained with the above
algorithm is the collected primary decomposition associated with the
Cartan subalgebra ${\cal A}$.

\bpr (Proof of 1, 2 and 3 in definition \ref{CPD}).  Condition 1 is
verified since if $X_2 \in {\cal A}$, and $V \in {\cal V}_j$, we
have (since $X$ and $X_2$ commute and therefore so do $ad_X$ and
$ad_{X_2}$) $ (ad_X^2+a_j^2{\bf 1})ad_{X_2}V=ad_{X_2}
(ad_X^2+a_j^2{\bf 1})V=0.  $ Therefore $ad_{X_2} V \in {\cal V}_j$
as well.  Condition 2 is also verified. Since ${\cal V}_j$ is two
dimensional, the minimum polynomial of $ad_{X_2}$ restricted to
${\cal V}_j$, for every $X_2 \in {\cal A}$, must be of the form
$\lambda^2+b^2$ for real $b$. Any other form would imply that
$ad_{X_2}$ has eigenvalues with nonzero real part, which has to be
excluded because of Proposition \ref{Richard} in the Appendix.
Condition 3 is verified taking as element $X$ in ${\cal A}$
precisely the splitting element.

\epr

\subsection{Calculation of the decomposition in simple ideals}
\label{simid}

The primary decomposition is a fundamental tool to explore the
structure of a semisimple Lie algebra ${\cal S}$. Using it, one can
directly obtain the decomposition into simple ideals (\ref{ssdec}).
The  algorithm is given in \cite{DeGraaf} (cf., Section 4.12) and we
report it using our notations.

\vs

\verb"Algorithm 3"
\begin{enumerate}

\item For every $j=1, \ldots,r$, calculate the spaces
\be{calI} {\cal I}_j:=\bigoplus_{k=0}^\infty ad_{\cal S}^k {\cal
V}_j, \ee where ${\cal V}_j$ are defined in (\ref{ioplm}).

${\cal I}_j$ is the smallest ideal containing ${\cal V}_j$.
 $ad_{\cal S}^k {\cal
V}_j,$ is defined inductively, where $ad_{\cal S}^0 {\cal V}_j={\cal
V}_j$, and $ad_{\cal S}^k{\cal V}_j=[{\cal S}, ad_{\cal
S}^{k-1}{\cal V}_j]$.

\item The simple ideals ${\cal S}_l$, $l=1,\ldots,p$ in
(\ref{ssdec}) are given by the ideals ${\cal I}_j$. Notice that some
ideals may be coinciding.
\end{enumerate}

\vs

\bpr (Proof of Algorithm 3) The main fact used to justify the
algorithm is that the primary decomposition (\ref{ioplm}) is
compatible with the decomposition in simple ideals (\ref{ssdec}).
This is proved in \cite{DeGraaf} (Theorem 4.12.1) and it means the
following: The Cartan subalgebra of $\cal S$, ${\cal A}$,  splits in
$p$ subalgebras, ${\cal A}_j$, $j=1,\ldots,p$, i.e., $ {\cal
A}=\bigoplus_{j=1}^p {\cal A}_j, $ where each ${\cal A}_j$ is a
Cartan subalgebra of the corresponding ${\cal S}_j$. Moreover, each
${\cal V}_j$, $j=1,\ldots,r$,   in (\ref{ioplm}) is a subspace of
one of the ${\cal S}_k$, $k=1,\ldots,p$. From the latter fact, it
follows immediately that each ${\cal S}_k$, $k=1,\ldots,p$, is the
smallest ideal generated by one ${\cal I}_j$, that is, it is of the
form (\ref{calI}). In fact, each ${\cal I}_j$ is contained in some
${\cal S}_k$ and ${\cal S}_k$ cannot contain any nontrivial ideal
other than itself, being simple. \epr

\section{Example: Control of two interacting spin $\frac{1}{2}$'s}
\label{examples}

Consider the control of two interacting spin $\frac{1}{2}$ particles
subject to a magnetic field. The state of particle $1$ ($2$) lives
 in a 2-dimensional Hilbert space ${\cal H}_1$ (${\cal H}_2$), so
 that the state of the total system lives in a $4$-dimensional
Hilbert space ${\cal H}_1 \otimes {\cal H}_2$. This type of systems
are of interest for example in quantum computation when one wants to
perform  operations with two quantum bits (cf., e.g., \cite{NC}). In
the model we shall consider an externally  applied magnetic field is
constant, has  nonzero component in the $x$ direction only and it
only affects the first spin. It is however possible to control the
interaction between the two spins. There are several ways to
experimentally achieve this; see, e.g., \cite{Kane}. The Hamiltonian
$H(u)$ in the system's equation (\ref{Mgensys}) has the form
\be{formaHam} H=u_1(t) \sigma_z \otimes \sigma_z + u_2(t) \sigma_y
\otimes \sigma_y + \sigma_x \otimes {\bf 1}   \ee where
$\sigma_{x,y,z}$ are the Pauli matrices
$\sigma_x:=\frac{1}{2}\pmatrix{0 & 1 \cr 1 & 0},$ $
\sigma_y:=\frac{1}{2}\pmatrix{0 & i \cr -i & 0},$
$\sigma_z:=\frac{1}{2}\pmatrix{1 & 0 \cr 0 & -1},$ which satisfy the
commutation relations \be{commurel} [i\sigma_x, i \sigma_y]=i
\sigma_z, \qquad [i \sigma_y, i \sigma_z]=i \sigma_x, \qquad [i
\sigma_z, i\sigma_x]= i \sigma_y.  \ee The solution $X$ of
(\ref{Mgensys}) represents the evolution on ${\cal H}_1 \otimes
{\cal H}_2$ as the state $\psi(t)$ evolves as $\psi(t)=X(t) \psi(0)$
with $X(t)$ the solution of (\ref{Mgensys}). The dynamical Lie
algebra is generated by $\{ i \sigma_z \otimes \sigma_z, i \sigma_y
\otimes \sigma_y, i \sigma_x \otimes {\bf 1} \}$. It is given by
\be{dynamLiealg} {\cal L}:=span \{i \sigma_x \otimes {\bf 1}, i {\bf
1} \otimes \sigma_x, i \sigma_z \otimes \sigma_z, i \sigma_y \otimes
\sigma_y, i \sigma_z \otimes \sigma_y, i \sigma_y \otimes \sigma_z
\},  \ee which is $6$-dimensional. As the dimension of the full Lie
algebra $su(4)$ is $15$, this shows that the system is not
controllable. The Lie group $e^{\cal L}$ gives the set of reachable
evolutions. In order to understand the nature of this set and
perform control to any possible value in it, we apply the analysis
developed in this paper. An application of the algorithm for the
calculation of the (Abelian) radical $\cal R$ discussed at the
beginning of Section \ref{algo} shows that ${\cal R}=\{ 0\}$ and
that ${\cal L}$ is semisimple. To calculate the simple subalgebras,
we apply the algorithms developed in Section \ref{algo}. We apply
\verb"Algorithm 1" to find the Cartan subalgebra ${\cal A}$.
Selecting $X=i \sigma_x \otimes {\bf 1}$ at Step 2 of that
algorithm, we find \be{Cartansub} {\cal A}=span \{i \sigma_x \otimes
{\bf 1}, i {\bf 1} \otimes \sigma_x \}.  \ee To find the primary
decomposition according to \verb"Algorithm 2" we have to select a
splitting element in ${\cal A}$. We write the adjoint
representations of $i\sigma_x \otimes {\bf 1}$ and $i {\bf 1}
\otimes \sigma_x$ in the ordered basis indicated in
(\ref{dynamLiealg}). We have $$ ad_{i \sigma_x \otimes {\bf
1}}=\pmatrix{0 & 0 & 0 & 0 &0 &0 \cr 0 & 0 & 0 & 0 &0 &0 \cr 0 & 0 &
0 & 0 &0 &1 \cr 0 & 0 & 0 & 0 &-1 &0 \cr 0 & 0 & 0 & 1 &0 &0 \cr 0 &
0 & -1 & 0 &0 &0}, \quad ad_{i {\bf 1} \otimes \sigma_x}=\pmatrix{0
& 0 & 0 & 0 &0 &0 \cr 0 & 0 & 0 & 0 &0 &0 \cr 0 & 0 & 0 & 0 &1 &0
\cr 0 & 0 & 0 & 0 &0 &-1 \cr 0 & 0 & -1 & 0 &0 &0 \cr 0 & 0 & 0 & 1
&0 &0}. $$ Neither $i \sigma_x \otimes {\bf 1}$ nor $i{\bf 1}
\otimes {\sigma_x}$ is a splitting element but $i \sigma_x \otimes
{\bf 1} +2 i{\bf 1} \otimes {\sigma_x}$ is because $ad_{i \sigma_x
\otimes {\bf 1} +2 i{\bf 1} \otimes {\sigma_x}}=ad_{i\sigma_x
\otimes {\bf 1}}+2 ad_{i {\bf 1} \otimes {\sigma_x}}$ has $\dim
{\cal L}- \dim {\cal A}+1=6-2+1=5$ different eigenvalues, $0, \pm
3i, \pm i$. The eigenspaces corresponding to $\pm 3i$ and $\pm 1$
are given by $ {\cal V}_1= span \{ i (\sigma_z \otimes \sigma_y +
\sigma_y \otimes \sigma_z), i (\sigma_z \otimes \sigma_z - \sigma_y
\otimes \sigma_y)\},   $ and $ {\cal V}_2=span \{ i (\sigma_z
\otimes \sigma_y - \sigma_y \otimes \sigma_z), i (\sigma_z \otimes
\sigma_z + \sigma_y \otimes \sigma_y)\}.  $ With ${\cal A}$ in
(\ref{Cartansub}) and ${{\cal V}_1}$ and ${\cal V}_2$ above, the
primary decomposition is ${\cal L}={\cal A} \oplus {\cal V}_1 \oplus
{\cal V}_2. $ Using \verb"Algorithm 3", we obtain the simple
component Lie algebras ${\cal S}_A$, ${\cal S}_B$, as the ideals
generated by ${\cal V}_1$ and ${\cal V}_2$. We have $ {\cal L}={\cal
S}_A \oplus {\cal S}_B, \qquad [{\cal S}_A, {\cal S}_B]=0, $ with
the simple Lie algebras ${\cal S}_A$ and ${\cal S}_B$ given by (cf.
(\ref{calI})) ${\cal S}_A=\bigoplus_{k=0}^\infty ad_{\cal L}^k {\cal
V}_1:=span \{A_1,A_2,A_3\}, \qquad {\cal S}_B=\bigoplus_{k=0}^\infty
ad_{\cal L}^k {\cal V}_2:= span \{B_1,B_2,B_3\}, $ with $
A_1:=\frac{i}{2}(\sigma_x \otimes {\bf 1}+{\bf 1}\otimes
{\sigma_x}),$ $A_2:=\frac{i}{2} (\sigma_z \otimes \sigma_y +
\sigma_y \otimes \sigma_z),$ $ A_3:=\frac{i}{2}(\sigma_z \otimes
\sigma_z-\sigma_y\otimes \sigma_y), $ and
$B_1:=\frac{i}{2}(-\sigma_x \otimes {\bf 1}+{\bf 1} \otimes
\sigma_x),$ $B_2:=\frac{i}{2} (\sigma_z \otimes \sigma_y - \sigma_y
\otimes \sigma_z),$ $ B_3:=\frac{i}{2}(\sigma_z \otimes
\sigma_z+\sigma_y\otimes \sigma_y). $ The commutation relations
$[A_1,A_2]=A_3,$ $[A_2,A_3]=A_1$, $[A_3,A_1]=A_2$ ($[B_1,B_2]=B_3,$
$[B_2,B_3]=B_1$, $[B_3,B_1]=B_2$), compared with (\ref{commurel}),
show that ${\cal S}_A$ and ${\cal S}_B$ are both isomorphic to
$su(2)$. Writing the Hamiltonian (\ref{formaHam}) as $
iH=iH_A+iH_B,$ $iH_A=A_3(u_1-u_2)+A_1,$  $ iH_B=B_3(u_1+u_2)-B_1,$
we find that the solution $X$ of (\ref{Mgensys}), (\ref{formaHam})
is the commuting product of the solutions $U_A$ and $U_B$ of the
decoupled systems $\dot U_A=-iH_A(u_1-u_2)U_A, \qquad U_A(0)={\bf
1}, $ and $ \dot U_B=-iH_B(u_1+u_2)U_B, \qquad U_B(0)={\bf 1},
 $i.e., $X(t)=U_A (t) U_B (t)=U_B(t)U_A(t)$. We can use $u_1-u_2$ and
$u_1+u_2$ as independent controls to drive $U_A$ and $U_B$,
respectively. Because of the isomorphism between ${\cal S}_A$ and
${\cal S}_B$ with $su(2)$, both control problems are equivalent to
control problems on $SU(2)$ for which there exists a large
literature. One can for example use the Riemannian symmetric space
argument of \cite{Khancool} to obtain the minimum (in fact infimum)
time control if there is no bound on the control. Alternatively one
can use an optimal, minimum energy, control over a finite time
horizon which turns out to be given by elliptic functions as
described in \cite{conmoh} or a Lie group decomposition technique as
in \cite{IoAutom} which can be applied when there are bounds on the
magnitude of the controls.





\section*{Appendix: Some facts about Lie algebras and Lie groups}

A {\bf Lie algebra} (see e.g. the textbooks  \cite{DeGraaf},
\cite{Humphreys}, \cite{SagleWalde}) is a vector space closed under
with a binary operation called the commutator $(x,y) \rightarrow
[x,y]$ which is bilinear with respect to the vector space sum,
skew-symmetric ($[x,y]=-[y,x]$) and satisfies the {\it Jacobi
identity}: $[x,[y,z]]+[y,[z,x]]+[z,[x,y]]=0$. For Lie algebras of
matrices that are the ones that interest us in this paper, the
commutator is taken to be the usual commutator of two matrices
$[A,B]:=AB-BA$. The Lie algebra $u(n)$ ($su(n)$) is the Lie algebra
over the reals of $n\times n$ skew-Hermitian matrices (with trace
zero). It has dimension $n^2$ ($n^2-1$). Two Lie algebras ${\cal L}$
and ${\cal L}'$ are {\bf isomorphic}, if there exists a linear one
to one and onto map $\phi: {\cal L} \rightarrow {\cal L}'$, such
that $\phi([A,B]_{\cal L})=[\phi(A),\phi(B)]_{{\cal L}'}$, for every
$A$ and $B$ in ${\cal L}$. Here $[ \cdot , \cdot ]_{\cal L}$ and $[
\cdot , \cdot ]_{{\cal L}'}$ denote the commutators in ${\cal L}$
and ${\cal L}'$ respectively. Given a Lie algebra ${\cal L}$ over
the reals, it is possible to define a Lie algebra over the complex
field ${\cal L}_{\otimes \CC}$ which is called the {\bf field
extension} of ${\cal L}$. ${\cal L}_{\otimes \CC}$  has the same
basis as ${\cal L}$ and it has the same dimension over the complex
numbers as ${\cal L}$ over the reals. The Lie brackets between two
basis elements give the same result as for the real Lie brackets .
Associated with a Lie algebra ${\cal L}$ of matrices is a {\bf Lie
group} $e^{\cal L}$ which is defined as the set of finite products
of the form $e^{L_1}e^{L_2}\cdot \cdot \cdot e^{L_f}$, with
$L_1,\ldots,L_f \in {\cal L}$, for $f \geq 0$. A Lie group is a
group in the algebraic sense and it is a differentiable manifold.
Naturally, the open sets in $e^{\cal L}$ are defined by requiring
that the map $\pi: \RR^n \rightarrow e^{\cal L}$,
$\pi(t_1,\ldots,t_n)=e^{A_1 t_1}\cdots e^{A_n t_n}$, for any
$\{A_1,\ldots A_n \}$ basis in $\cal L$ is open, i.e., maps open
sets in $\RR^n$ into open sets in $e^{\cal L} $. The Lie group
associated with the Lie algebra $u(n)$ ($su(n)$) is the Lie group of
$n \times n$ unitary matrices $U(n)$ ($n \times n$ unitary matrices
with determinant equal to $1$, $SU(n)$). A {\bf Lie subgroup} $\bf
S$ of $e^{\cal L}$ is a subgroup in the algebraic which is also a
sub-manifold in the sense that the topology of $\bf S$ coincides
with the one induced by the one of $e^{\cal L}$. If ${\cal K}$ is a
subalgebra of $\cal L$, then the Lie group $e^{\cal K}$ is a Lie
subgroup of $e^{\cal L}$ provided that the last condition on the
topology is satisfied.\footnote{This is tacitly assumed anytime we
talk about a Lie subgroup in the paper. The result mentioned at the
beginning of the introduction has to be slightly modified if this
last assumption is not verified by saying that the set of reachable
values of $X$ for (\ref{Mgensys}) is {\it dense} (in the topology of
$U(n)$) in the Lie group associated to the dynamical Lie algebra
${\cal L}$. Therefore from a practical point of view there is no
difference whether $e^{\cal L}$ is or is not a Lie subgroup of
$U(n)$. The author wishes to thank Francesco Ticozzi for useful
discussions on this point.} Consider a Lie algebra ${\cal L}$, and
define inductively the following sequence of subalgebras ${\cal
L}^0:={\cal L}$, ${\cal L}^{(k+1)}=[{\cal L}^{(k)}, {\cal
L}^{(k)}]$. ${\cal L}$ is called {\bf solvable} if there exists a
$k$ such that ${\cal L}^{(k)}=\{0\}$. Another sequence is given by
${\cal L}_0={\cal L}$, ${\cal L}_{k+1}=[{\cal L}, {\cal L}_k]$. A
Lie algebra is {\bf nilpotent} if, there exists a $k$ such that
${\cal L}_k=\{0\}$. It is called {\bf Abelian} if ${\cal
L}_1=\{0\}$. If a Lie algebra is Abelian it is nilpotent. If it is
nilpotent it is solvable. An {\bf ideal} of ${\cal L}$ is a subspace
${\cal I}$, such that $[{\cal L}, {\cal I}] \subseteq {\cal I}$. A
Lie algebra ${\cal L}$ is called {\bf simple} if it has dimension
$>1$ and it contains no ideals except the trivial ones, $\{ 0 \}$
and ${\cal L}$. A Lie algebra is {\bf semisimple} if it is the
direct sum (i.e., the sum of vector spaces which commute with each
other) of simple Lie algebras.
 Semisimple Lie algebras ${\cal S}$ have the property that $[{\cal
 S} , {\cal S}]={\cal S}$. Consider a subspace ${\cal N}$ of the Lie
 algebra ${\cal L}$. The {\bf normalizer} of ${\cal N}$, is the set
 $N_{\cal L}({\cal N})=\{L \in {\cal L}|[L,{\cal N}]\subseteq {\cal
 N}\}$. The {\bf center} of a Lie algebra
 ${\cal L}$ is the set $C({\cal L}):=\{L \in {\cal L}|[L,{\cal L}]=0\}$.
 It is clear that the center is a subspace of the normalizer and
 both of them (using Jacobi identity) are  closed under commutation
 and therefore are subalgebras of ${\cal L}$. Let ${\cal L}$ be a Lie algebra over the field of reals.  The {\bf
adjoint representation} of the Lie algebra ${\cal L}$ is a function
$ad: {\cal L} \rightarrow {\cal M}_{\dim ({\cal L}), \dim ({\cal
L})}$\footnote{${\cal M}_{n,n}$ denotes the Lie algebra of $n\times
n$ real matrices with the commutator given by the standard matrix
commutator.} which maps $X\in {\cal L}$ to  $ad_X$, where $ad_X$ is
a linear map ${\cal L} \rightarrow {\cal L}$ defined by $ad_X
L:=[X,L]$. The adjoint representation is a {\bf representation} of
the Lie algebra ${\cal L}$ in that it preserves the basic Lie
algebra operations.  We have $ad_{[X,Y]}=[ad_X,ad_Y]$ and  $ad_X$ is
a matrix acting on $\RR^{\dim {\cal L}}$. The following fact is used
in the paper.  \bp{Richard} If ${\cal L}$ is semisimple, for every
$X \in {\cal L}$, there exists a basis in ${\cal L}$ such that
$ad_X^T=-ad_X$.\footnote{The author thanks Richard Ng for working
out this proof and for helpful discussions on this paper.} \ep \bpr
Consider the {\it Killing form} on ${\cal L}$, $\langle Y, Z
\rangle_K$ defined as $\langle Y, Z \rangle_K:= Tr (ad_Y ad_Z).  $
Since ${\cal L}$ is semisimple according to Cartan criterion (see,
e.g., \cite{DeGraaf}) the Killing form is non-degenerate. Consider a
basis of ${\cal L}$, which is orthogonal with respect to the Killing
form. If we define the transposed of a linear operator $O$ by
$\langle Y, O^T(Z) \rangle_K=\langle O Y, (Z) \rangle_K$, it is
easily seen that the matrix form of the transposed in the
orthonormal basis is the usual transposed of the  matrix form of
$O$.  We have, given $X$, for every $Y$ and $Z$, $ \langle Y,
ad_{X}^TZ \rangle_K:=\langle ad_X Y,Z\rangle_K= \langle [X,Y], Z
\rangle_K= \langle [Z,X], Y \rangle_K = \langle Y, [Z,X] \rangle_K
=- \langle Y, ad_X Z \rangle_K. $ The first equality follows from
the definition of transposed. Then, we used the cyclic property of
the Killing form in the third equality and its symmetry in the
fourth equality. Summarizing, we have $ \langle Y, ad_X^TZ
\rangle_K= - \langle Y, ad_XZ \rangle_K.  $ Since the Killing form
is non-degenerate, we must have $ad_X^TZ=-ad_XZ$ and since this is
true for every $Z$ the claim follows.  \epr

\end{document}